# Robust Luttinger liquid state of 1D Dirac fermions in a van der Waals system $Nb_9Si_4Te_{18}$

Qirong Yao[a], Hyunjin Jung[a,b], Kijeong Kong[a], Chandan De[a], Jaeyoung Kim[a], Jonathan D. Denlinger[c], and Han Woong Yeom[a,b]*

[a]*Center for Artificial Low Dimensional Electronic Systems, Institute for Basic Science (IBS), Pohang 37673, Korea*

[b]*Department of Physics, Pohang University of Science and Technology, Pohang 37673, Korea*

[c]*Advanced Light Source, Lawrence Berkeley National Laboratory, CA 94720, USA*

*Email: yeom@postech.ac.kr

**Abstract**

We report on the Tomonaga-Luttinger liquid (TLL) behavior in fully degenerate 1D Dirac fermions. A ternary van der Waals material $Nb_9Si_4Te_{18}$ incorporates in-plane $NbTe_2$ chains, which produce a 1D Dirac band crossing Fermi energy. Tunnelling conductance of electrons confined within $NbTe_2$ chains is found to be substantially suppressed at Fermi energy, which follows a power law with a universal temperature scaling, hallmarking a TLL state. The obtained Luttinger parameter of ~0.15 indicates strong electron-electron interaction. The TLL behavior is found to be robust against atomic-scale defects, which might be related to the Dirac electron nature. These findings, as combined with the tunability of the compound and the merit of a van der Waals material, offer a robust, tunable, and integrable platform to exploit non-Fermi liquid physics.

When electrons are confined into one dimension, their mutual interaction is substantially enhanced and predicted to drive the system out of the Fermi-liquid state into separate collective excitations of spins and charges as pictured by Tomonaga-Luttinger-liquid (TLL) theory.[1-10] Those collective excitations follow a power-law decay of spin and charge correlation functions, which in turn govern various low-energy physical properties. Spectroscopically, such TLL physics is manifested by the power-law suppression of density of states (DOS) at Fermi energy as observed in quasi-one-dimensional (quasi-1D) crystals,[11] carbon nanotubes,[12] quantum wires,[13] quantum wells,[14,15] atom chains,[16] nanoribbons,[17] grain boundaries or edges of 2D crystals[18-22], and in organic crystals.[23,24]

While most of TLL behaviors are observed for systems with non-relativistic electronic bands, consistent behaviors were also reported for rare 1D Dirac bands such as spin-polarized Dirac bands of quantum spin Hall edge states and fully generate Dirac bands of carbon nanotubes.[25-27] Since the TLL behavior has been discussed in terms of the linearized-band dispersion within a very narrow energy window near Fermi energy, the substantial difference of a Dirac 1D system is not straightforwardly expected. However, the possible consequence of the time-reversal symmetry of a 1D Dirac band and the Dirac point in the TLL behavior has not been made sufficiently clear. For example, TLL behaviors observed so far in non Dirac bands have been known to be very vulnerable to extrinsic defects and impurities due intrinsically to the strong backscattering of 1D electrons by defects. This has been an important limiting factor in preparing a system with well-defined TLL behaviors. In contrast, the robust TLL behavior against structural defects (kinks) was reported recently for the spin-polarized helical Dirac bands of quantum spin Hall edge states, indicating the potential importance of 1D Dirac electrons in exploiting TLL physics.[25]

In the present work, we establish a van der Waals material $Nb_9Si_4Te_{18}$ as a TLL system and reveal its robustness of the TLL behavior against impurities. $Nb_9Si_4Te_{18}$ belongs to one member of composition-tunable ternary compounds $Nb_{2n+1}Si_nTe_{4n+2}$ with n=4, which was recently shown to host fully degenerate spinful 1D Dirac bands. Here, we have analyzed the low energy tunneling spectra near the Fermi level of $Nb_9Si_4Te_{18}$. It turns out that the low energy spectral intensity is strongly suppressed following well a universal power-law scaling behavior in both energy and temperature, indicating unambiguously a TLL state. Furthermore, this TLL behavior is found to be robust against impurities, which is possibly related to the distinct backscattering property of Dirac electrons irrespective of its topological nature. The present findings together with the tunability of the compounds and the merit of the van der Waals material architecture make this material an attractive platform for studying non-Fermi liquid physics.

The monolayer structure model of the $Nb_9Si_4Te_{18}$ crystal is displayed in Figure 1a, where Nb and Si atoms are located at the same plane as sandwiched between two Te layers. The in-plane crystal structure of the compound can be viewed as an alternation of a $NbTe_2$ chain (marked by pink bars) and a bundle of four $Nb_2SiTe_4$ chains so that its chemical formula can be rewritten as $(NbTe_2)(Nb_2SiTe_4)_4$. A highly resolved STM topographic image of a $Nb_9Si_4Te_{18}$ surface is shown in Figure 1b. This relatively large negative sample bias suppresses possible electronic effects so that all the top-layer Te atoms (orange balls in the inset of Figure 1b) are clearly resolved. One can easily distinguish two types of 1D structures, which correspond to $NbTe_2$ chains (black dashed box) and $(Nb_2SiTe_4)_4$ bundles (white dashed box), respectively. The band structure of a monolayer $Nb_9Si_4Te_{18}$ calculated by density functional theory (DFT) is presented in Figure 1c. Its low-energy electronic states are dominated by a flat band along X-U direction (orthogonal to NbTe$_2$ chains), which disperses one-dimensionally along the chains to cross the Fermi energy. This 1D band has a fourfold degeneracy and Dirac crossings at the Brillouin zone boundary of X and U, which are protected by the nonsymmorphic glide mirror symmetry. Angle-resolved photoemission spectroscopy (ARPES) observes the zone boundary Dirac crossing at ~50 meV (X point) below the Fermi energy (Figure 1f). Its strong 1D character is confirmed by its straight traces in the Fermi surface map (Figure 1d) and the lack of dispersion as a function of photon energy (Figure 1e). As reported previously and predicted in the calculation, there also exists a 2D hole band at Γ points to result in Fermi surfaces of overlapped rings (Figure 1d-e).[28] In the calculation, while overall band dispersions are very similar for both monolayer and bulk,[29] a small splitting at the Dirac point is expected in the bulk. However, the present and the previous ARPES experiments do not find such a splitting.[28,29] The preserved Dirac crossing may be due to the lack of the bulk interlayer coupling in the surface layer probed or the limited resolution of the experiments.

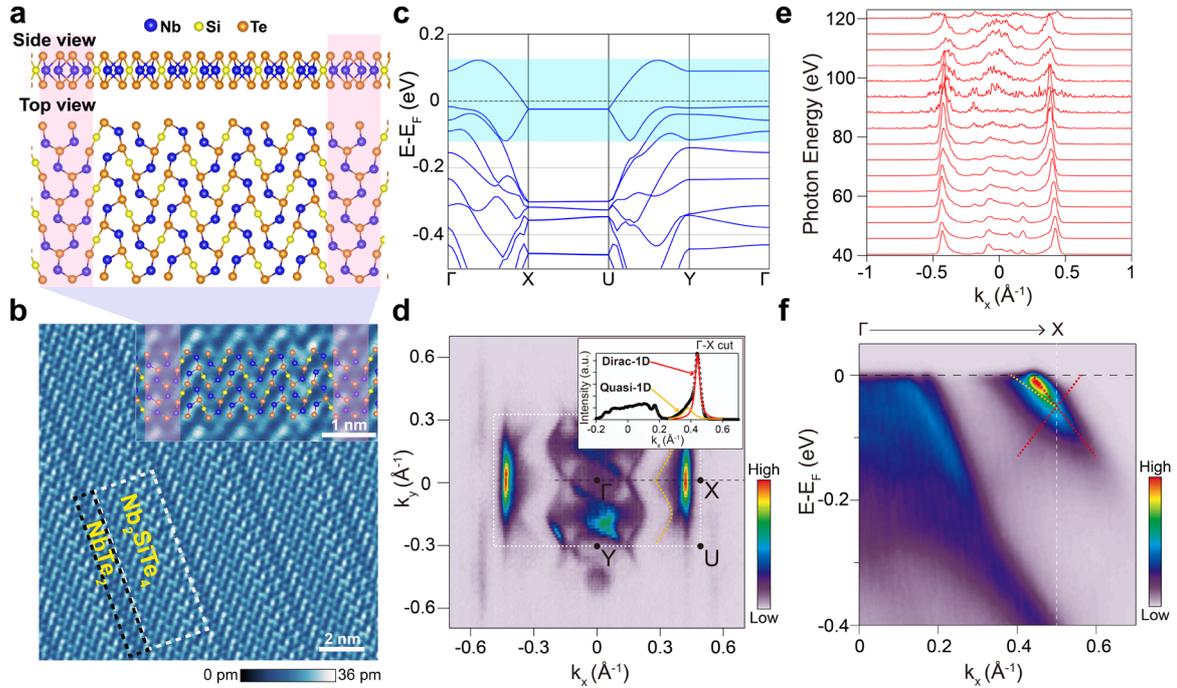

**Figure 1.** Surface morphology and band structure of $Nb_9Si_4Te_{18}$ crystal. a, Ball-and-stick model of a $Nb_9Si_4Te_{18}$ monolayer with top and side views, respectively. b, STM topographic image of the top surface with atomic resolution (scanning size 15 nm × 15 nm, V = -1 V, I = 50 pA). Inset: Zoomed-in STM image, revealing the positions of all Te atoms on the top surface. c, Theoretical calculated band structure of monolayer $Nb_9Si_4Te_{18}$. d, Constant energy contours of ARPES spectra near $E_F$, where the quasi-1D band is marked by the orange dashed line. e, Photon-energy dependent momentum distribution curves (MDCs) at the Fermi level along Γ-X direction. f, Electronic band map of ARPES along high-symmetric points at Γ-X direction.

The 1D confinement of the Dirac electrons, which comes from Nb 4d electrons along $NbTe_2$ chains in the DFT calculation, can further be revealed by STM and STS measurements. STM topography observes a bright 1D chain array with an interchain distance of 3.5 nm (Figure 2a). Its topographic contrast has a strong bias dependence; bright (dark) in the empty state of 0.4 V in Fig. 2a (the low filled-state bias) (see Supplemental Material, Figure S1 for a full bias set of STM images). This is consistent with the previous work[29] and the bright chains in the empty state can be identified as the $NbTe_2$ chains from their width and separation (see Figure 1b). Figure 2b shows a differential tunneling conductance (dI/dV), which is proportional to local density of states (LDOS), as a function of energy and distance perpendicular to 1D chains. The electrons of a $NbTe_2$ chain are strongly confined in 1D as manifested by their well-confined differential conductance signal within the chains, especially in the empty state. A strong LDOS peak is noticed at 0.15 eV in the empty state (-0.1 eV in the filled state),[29,30] which is due to the top (bottom) of the 1D Dirac band on the $NbTe_2$ chain. The 1D confinement was previously demonstrated in a different aspect, that is, by the formation of 1D quantum well states between neighboring defects along $NbTe_2$ chains.[29]

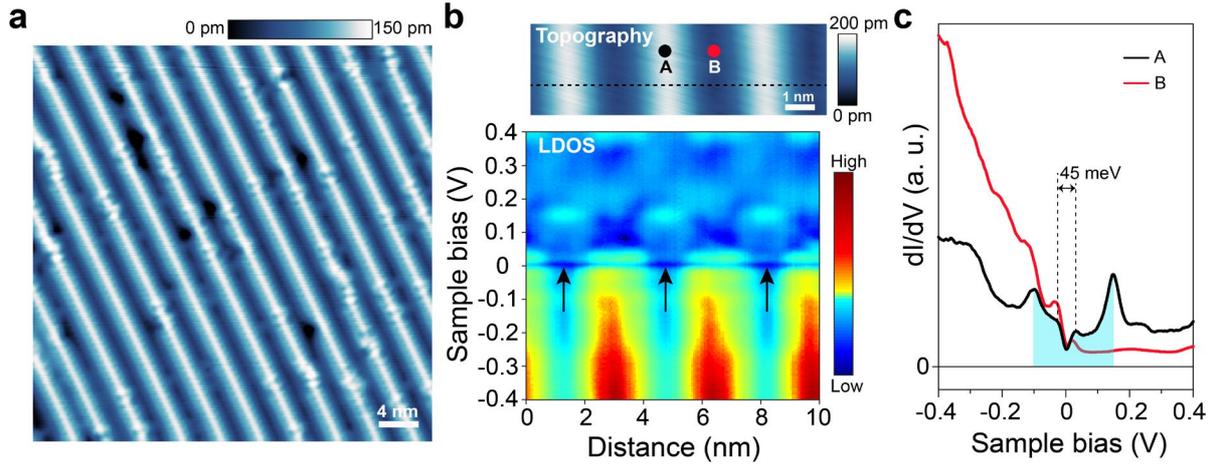

**Figure 2.** Suppression of the DOS near Fermi energy level in 1D NbTe$_2$ chain. a, STM topographic image of the top surface (scanning size 40 nm × 40 nm, V = 0.4 V, I = 50 pA). b, Line-scan dI/dV map taken along the black dashed line for a span of 10 nm, A corresponding to a NbTe$_2$ chain and B for (Nb$_2$SiTe$_4$)$_4$ bundle. c, Differential conductance dI/dV curves measured on the NbTe$_2$ chain and the (Nb$_2$SiTe$_4$)$_4$ bundle, displayed by black and red solid lines, respectively.

Beyond the 1D confinement, we note an unusual point of the spectra, namely, the strong suppression of the LDOS right at Fermi energy within an energy window of about 45 meV. As clearly shown in Figure 2b, this suppression is marginal on the other part of the surface but is enhanced on the 1D Dirac electrons of NbTe$_2$ chains (see the arrows). This can be more clearly shown in the comparison of the averaged dI/dV curves of NbTe$_2$ (black) chains and (Nb$_2$SiTe$_4$)$_4$ bundles (red) in Figure 2c. This suppression cannot be attributed to a band gap since there is a finite LDOS at the Fermi energy and the spectra around Fermi energy is a 'V' shape even at a temperature of as low as 1.1 K (Figure 3a). We also confirm that the suppression has no magnetic field dependence up to 2 T (Figure S2), ruling out the possibility of the superconducting fluctuation.

From the 1D band dispersion of electrons on NbTe$_2$ chains, we naturally check the TLL origin of the suppressed LDOS. Figure 3a shows the power-law ($\propto |E|^\alpha$) fit of STS spectra within the energy window of -20 to 20 meV centered at the Fermi level. The fit is good with a power-law exponent α =1.25 except at the Fermi energy. In fact, a universal power-law scaling in both energy and temperature can describe more details of TLL, which is described by the formula[26]

$$\rho(E,T) \propto T^\alpha \cosh\left(\frac{E}{2k_B T}\right) \left|\Gamma\left(\frac{\alpha+1}{2} + \frac{iE}{2\pi k_B T}\right)\right|^2$$

Here, $\rho(E,T)$ is the tunneling DOS of a TLL and Γ is the gamma function. It turns out that the STS spectrum at 1.1 K is in excellent agreement with the TLL model (the solid line in Figure 3a). Furthermore, all dI/dV curves between 4.4 and 40 K are fitted well with a consistent fitting parameter (α =1.27±0.08) as plotted in Figure 3d. The characteristic property of the universal scaling of TLL in energy and temperature can be confirmed by plotting the experimental dI/dV spectra scaled by $T^{1.27}$ against $E/k_B T$ (Figure 3c). Apparently, all spectra collapse into a single universal curve. This observation provides solid and direct evidence for the TLL nature of the 1D metallic state in NbTe$_2$ chains within a Nb$_9$Si$_4$Te$_{18}$ crystal.

There are a few possible scenarios that can result in the suppression of LDOS near the Fermi level.[16] Two major ones, the Coulomb blockade effect and the disorder effect can be ruled out immediately. The former is determined by the capacitance and the resistance of a tunneling junction[31] but the V-shaped spectra at Fermi energy are consistent over a large range of tunneling current from 10 pA to 2 nA (Figure S3). On the other hand, the STS spectra captured on defects exhibit a distinct behavior (Figure S3), which is easily distinguished from the TLL suppression.

The TLL theory tells that the power-law exponent α is directly related to the Luttinger parameter g, which characterizes the sign and the strength of the electron-electron interaction underlying the anomalous behavior. Regarding a standard spinful TLL, the relationship between α and g can be expressed by the equation

$$\alpha = (g + g^{-1} - 2)/4$$

With experimentally determined exponent α =1.27±0.08, the value of g is deduced to be ~0.15, departing well from the non-interacting Fermi liquid value of 1. This is comparable with that of a quasi-1D $Li_2Mo_6Se_6$ crystal[11] and establishes $Nb_9Si_4Te_{18}$ as a Dirac non-Fermi liquid system with strong electron-electron interaction. In a very recent STS study, the energy-dependent lifetime of holes along $NbTe_2$ chains in $Nb_7Si_3Te_{14}$ and $Nb_{11}Si_5Te_{22}$ was measured, which is consistent with the expectation of the TLL theory for $Nb_{11}Si_5Te_{22}$.[32] This is qualitatively consistent with the present direct measurements of the power law suppression of the DOS at the Fermi energy. This work further indicated that the power exponent is different between $Nb_7Si_3Te_{14}$ and $Nb_{11}Si_5Te_{22}$. While we do not find a substantial dependence of the power exponents for the interchain spacing over 3 nm (Figure S4), a smaller interchain distance (for example, $Nb_5Si_2Te_{10}$) was shown to suppress the TLL behavior due to the development of a quasi 2D band dispersion (see our ARPES data in the Supplemental Material, Figures S5-S6 and the previous STS spectra for crystals with different interchain spacings[29]). A further study on the detailed dependence of the TLL behavior on the interchain spacing is thus desirable. A similar case of a non-Fermi liquid state in a 1D Dirac Fermion system was recently reported for the quantum spin Hall edge state with STS.[25] This system belongs to a different class of TLL, a helical TLL, due to the spin-momentum locking and the value of Luttinger parameter was ~0.42 indicating a smaller degree of electron-electron interaction.[25] On the other hand, a spinful 1D Dirac system was reported for a metallic single-wall carbon nanotube, which indicates an even less interaction with a large value of g~0.55.[27]

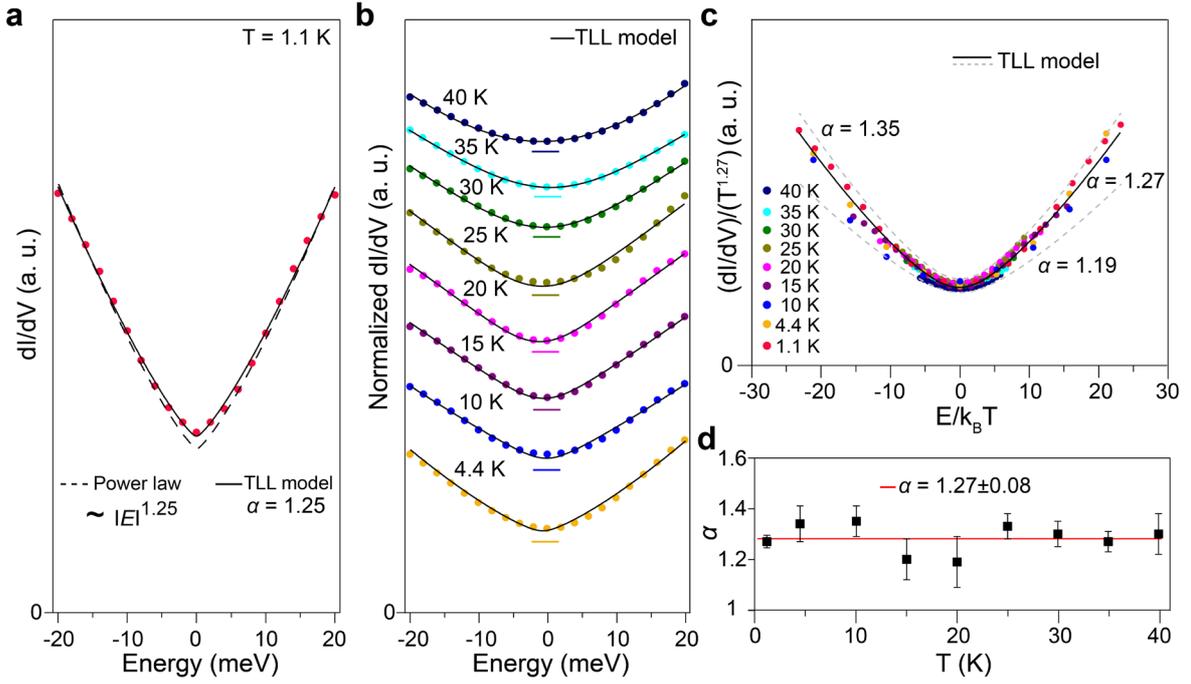

**Figure 3.** Universal scaling of power-law suppression of the DOS near the Fermi energy level. a, Differential tunneling conductance (dI/dV) curve taken on the 1D NbTe$_2$ chain at T= 1.1 K; the dashed line is the power-law fit while the black solid line corresponds to the fitting with equation (1), and the value of exponent α is 1.25. b, Temperature-dependent dI/dV curves taken on the same NbTe$_2$ chain in a, and the fitted curves with equation (1) are plotted by black solid lines. c, Scaled the dI/dV curves to $T^{1.27}$ as a function of temperature. d, Exponent α as a function of temperature, extracted from the TLL fittings in b, with an averaged value of α =1.27±0.08.

While the consequence of a Dirac dispersion in the TLL theory has not been investigated sufficiently, we suggest that the fundamental property of a Dirac band, the time-reversal symmetry, to suppress the backscattering[33-35] can influence interactions with defects. A STM topographic image for a short NbTe$_2$ segment is displayed in Figure 4a, which has a length of about 4.5 nm as terminated by two most popular point defects. The dI/dV map shows that the defects have noticeable LDOS only at a lower energy than -0.07 eV in the filled states and the quasiparticle interference is formed between the defects in agreement with the previous work.[29] However, it is notable that the LDOS suppression at the Fermi level is not significantly affected by the defects. The detailed dI/dV curves and the corresponding fittings indicate that the power-law behavior is surprisingly consistent to the very vicinity of the defect and deviates only near the defect site (due to its enhanced DOS at the filled states) within a length of about 0.4 nm (see Figure 4d and Supplemental Figure S7). This is clearly contrasted with a few previous TLL systems, where a significant difference of power-law behavior was observed in a long range from a defect.[16,36] Theoretically speaking, It is well established that even a weak impurity potential can be a perfect backscatter in TLL as the energy approaches the Fermi level to induce a pronounced and long-ranged influence on its power exponents.[37,38] We speculate that the the time-reversal symmetry of the Dirac band, which is protected by the nonsymmorphic symmetry, reduces backscattering as in the case of the Klein tunneling in graphene[35] to keep the TLL behavior robust. On the other hand, this weak defect perturbation may partly come from the lack of LDOS of the defect near Fermi energy or the tunneling process may wash the local variation under electron correlation. The atomistic origin of the defect is not clear at present while we suggest a Nb vacancy or a Nb intercalant (Figure S8). Further studies on the effect of various different defects would make important progress on the physics of TLL.

The power-law suppression would also be reflected in photoelectron spectra.[39-41] Our ARPES result shows clearly that the spectral function of the 1D Dirac band near the Fermi energy cannot be fitted by Fermi-Dirac function but follows a power-law behavior. However, a quantitative comparison of the power exponent between ARPES and STS spectra is found to be impractical due to the substantial overlap of the 1D Dirac band and an extra quasi-1D band (marked by the orange dashed line in Figure 1d and 1f) in the momentum space (Figure S5 and S6). This band was not mentioned in the previous ARPES work but one can notice its traces in the reported data therein.[28] In fact, this quasi-1D band dispersion (Figure S9) is consistent with those predicted for $Nb_5Si_2Te_{10}$ (Figure S10) indicating the coexistence of domains with different $NbTe_2$ chain spacings. Such lateral variation of the interchain spacing is indeed observed by STM (Supplemental Material, Figure S11 and Figure S12) as caused by the composition inhomogeneity. Nevertheless, a robust material platform of highly correlated 1D Dirac fermions within a van der Waals layer would provide further engineering opportunity, such as tuning of strain, doping, and interchain interaction, over similar states observed at edges or domain boundaries of 2D materials.

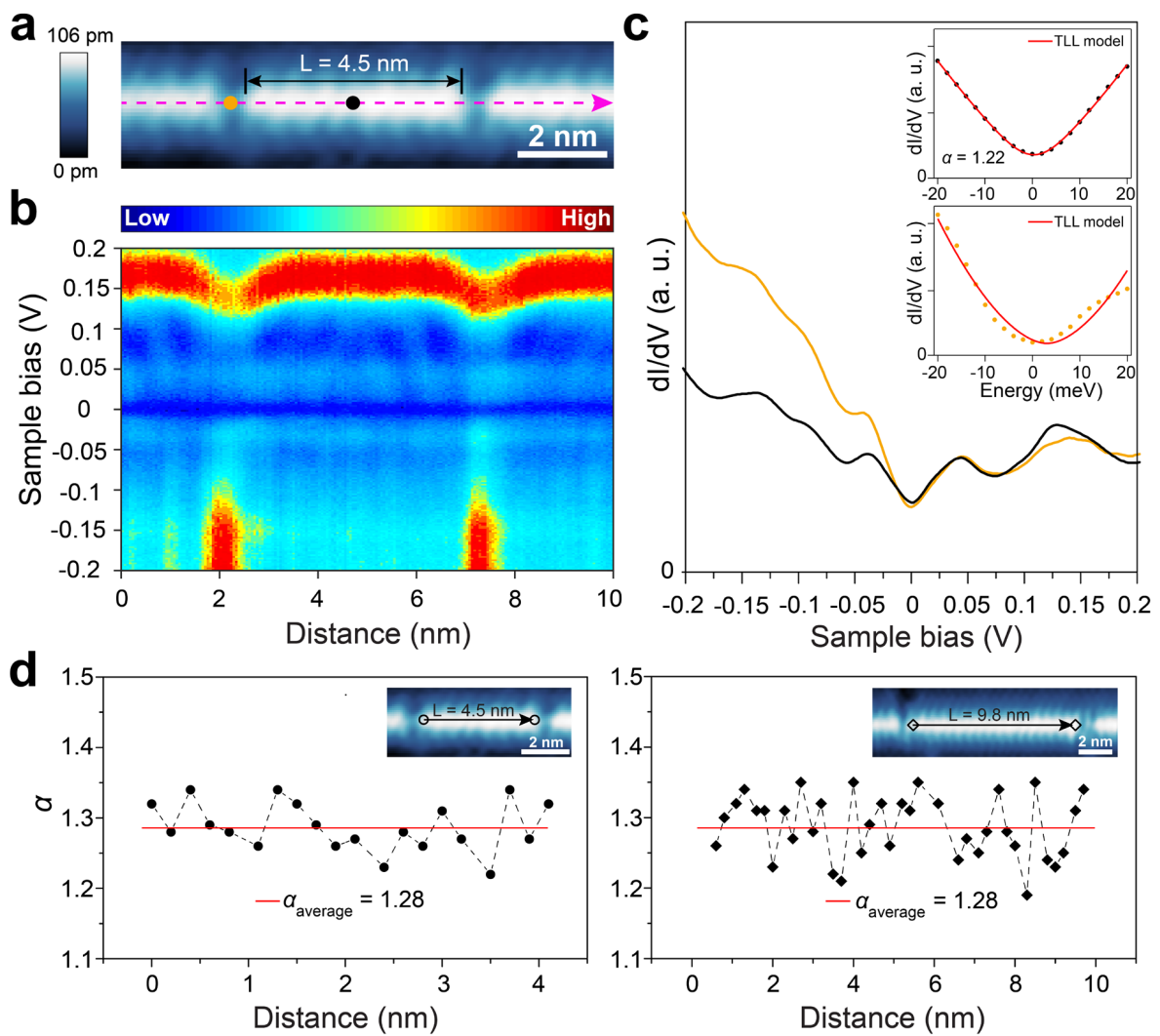

**Figure 4.** TLL behavior in short 1D $NbTe_2$ chains (length < 10 nm). a, STM topographic image of $NbTe_2$ chain with a length of 4.5 nm. b, Line-STS map captured along the pink dashed arrow in a. c, STS spectra taken on the spots in a, the black one is close to the center of $NbTe_2$ chain

while the orange one for the point-defect. Inset: power-law fitting based on the TLL model for both curves. d, Position-dependent exponent value extracted from the STS spectra taken on the $NbTe_2$ chain with lengths of 4.5 nm and 9.8 nm, respectively. Insert: STM topographic images of the two chains, and the STS spectra are captured along the black arrow marked on the chain.

To sum up, we have investigated the spectral anomaly of metallic 1D chains within a ternary van der Waals material $Nb_9Si_4Te_{18}$. The atom-resolved surface morphology and the electronic band structure are clearly revealed by STM and ARPES techniques, which confirm the existence of a 1D Dirac band along $NbTe_2$ chains within a van der Waals layer. The STS measurements for defect-free chains show the power-law suppression of the spectral weight at the Fermi energy following a universal temperature scaling, which hallmarks a TLL state. The Luttinger parameter extracted is 0.15, which indicates a strong deviation from the Fermi liquid and a strong electron correlation within the 1D Dirac band. The robust TLL behavior against point-defects is suggested to be originated from the reduced backscattering by the time-reversal symmetry of the Dirac band. This material platform of highly correlated 1D Dirac fermions within a van der Waals layer would provide unprecedented opportunity and controllability, such as tuning of strain and doping, over similar states observed at edges or domain boundaries of 2D materials.

# ASSOCIATED CONTENT

**Author information**

**Corresponding Authors**

**Han Woong Yeom** – *Center for Artificial Low Dimensional Electronic Systems, Institute for Basic Science (IBS), Pohang 37673, Republic of Korea; Department of Physics, Pohang University of Science and Technology, Pohang 37673, Republic of Korea*; orcid.org/0000-0002-8538-8993; Email: yeom@postech.ac.kr

**Author**

**Qirong Yao** - *Center for Artificial Low Dimensional Electronic Systems, Institute for Basic Science (IBS), Pohang 37673, Korea*

**Hyunjin Jung** - *Center for Artificial Low Dimensional Electronic Systems, Institute for Basic Science (IBS), Pohang 37673, Korea; Department of Physics, Pohang University of Science and Technology, Pohang 37673, Korea*

**Kijeong Kong** - *Center for Artificial Low Dimensional Electronic Systems, Institute for Basic Science (IBS), Pohang 37673, Korea*

**Chandan De** - *Center for Artificial Low Dimensional Electronic Systems, Institute for Basic Science (IBS), Pohang 37673, Korea*

**Jae Young Kim** - *Center for Artificial Low Dimensional Electronic Systems, Institute for Basic Science (IBS), Pohang 37673, Korea*

**Jonathan Denlinger** - *Advanced Light Source, Lawrence Berkeley National Laboratory, CA 94720, USA.*

**Author Contributions**

‡Q.Y. and H.J. contributed equally to this paper

Q.Y. performed STM experiments and analyzed the data. H.J., J.Y.K. and J.D. performed ARPES experiments and analyzed the data. K.K. performed DFT calculations. C.D. synthesized the crystals. H.W.Y. conceived the research idea and plan. Q.Y. and H.W.Y. wrote the paper with the inputs from all coauthors.

**Notes**

The author declare no competing financial interest.

**Supporting Information**

Experimental methods including crystal growth, STM & ARPES measurements, and DFT calculations. Bias-dependent STM topographic images, magnetic field-dependent dI/dV spectra, Tunneling current-dependent dI/dV spectra, STM/STS data analysis for the $NbTe_2$ chains, ARPES data analysis, bias-dependent STM topographic images with the point-defect, Theoretically calculated constant energy contours near the Fermi level depending on the Si concentration of $Nb_{2n+1}Si_nTe_{4n+2}$ family, STM topographic images with different interchain distances.

**ACKNOWLEDGMENT**

This work was supported by Institute for Basic Science (Grant No. IBS-R014-01).

TOC graphic

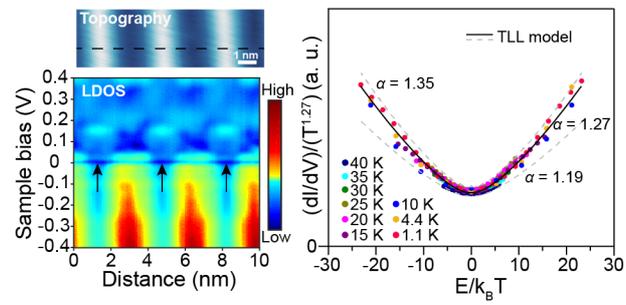